\documentclass[dvips]{article}
\usepackage{icrctc07}

\title{Gamma/hadron separation in IACTs using 3D EAS  variables}
\shorttitle{3D Gamma/Hadron separation}

\authors{S. ANDRINGA$^{1}$,P. ASSIS$^{1}$,M. PIMENTA$^{1}$,A. PINA$^{1}$,B. TOM\'E$^{1}$}
\shortauthors{Author and et al.}
\afiliations{$^1$LIP, Laborat\'orio de Instrumenta\c{c}\~ao e F\'isica Experimental de Part\'iculas, Lisbon, Portugal}
\email{pimenta@lip.pt}

\abstract{A new approach to Gamma/Hadron separation algorithms is proposed. The 
differences between Gamma and Hadron showers are notorious in two main aspects.
The first is the wideness of the shower, and the second is the distribution of the angles of emission 
of Cherenkov photons in respect to the shower main axis. Using more than one IAC 
telescope, and their respective bi-dimensional images of arrival directions of the 
Cherenkov photons, the 3D geometrical characteristics of the shower can be reconstructed.}

\begin{document}
\maketitle

\section{Introduction}
In recent years, the search for Gamma Ray sources has become one of the main concerns for astrophysics. 
The development of ground-based Imaging Atmospheric Cherenkov Telescopes (IACTs) allowed the detection of Gamma Ray showers
in the Very High Energy domain (in the order of the TeV) \cite{Cawley, Hillas}. This opened a possibility for new physics to be studied.

The number of known emitters has greatly increased in the last years. However, most of them remain unidentified and the 
prospect of discovering new objects is good. Several telescopes were developed to study these sources and, simultaneously, 
search for new ones.  

Most of the IACTs already built, as well as the ones planned for the near future, are characterized by having
a Field of View (FoV) of only a few degrees. While providing good resolutions for the study of the known sources, 
a sky survey of a considerable portion of the sky would consume too much time.

A new approach to the IAC technique is being proposed by GAW \cite{GAW}, that will introduce a large FoV up to 24$^{\circ}$x 24$^{\circ}$. 
One of the main goals in GAW is the serendipity search of sources, by performing a sky survey of 360$^{\circ}$x 60$^{\circ}$. 
In order to effectively locate sources, a good shower reconstruction algorithm has to be used. 
Furthermore, it is also necessary to distinguish the $\gamma$-ray showers from the ones induced by other primaries.

In GAW, three telescopes will be placed at the vertices of an aproximately equilateral triangle of 80 m side. This configuration was 
used as a case study to test the proposed method. The origin of the reference system is located at the centre of the triangle.

Generally, a bi-dimensional analysis of the images obtained by the IACTs is performed to calculate the Hillas parameters \cite{Hillas}. These 
are then used to characterize the detected shower and determine the type of primary that originated it. 
The Hillas parameters analysis is based solely on the projections of the shower in each telescope. Therefore, there is the
possibility that additional information can be used by reconstructing the shower in 3D.

With the positions of the telescopes and the arrival directions of the Cherenkov photons emitted in the cascade, a geometrical 
representation of the shower can be constructed. Different primaries have different characteristics, such as 
the shower being wider for proton induced showers than $\gamma$ induced ones. These differences can be 
translated in variables such as the impact parameter of each Cherenkov photon arrival direction ($b$), which is the minimum distance between 
the arrival direction and the shower main axis. The shower wideness also relates to differences in the distribution of $\alpha$, 
defined as the angle between the arriving photon direction and the main axis of the shower. A geometrical representation of both 
variables can be seen in figure 1.

\begin{figure}[h]
\begin{center}
\includegraphics [width=0.35\textwidth]{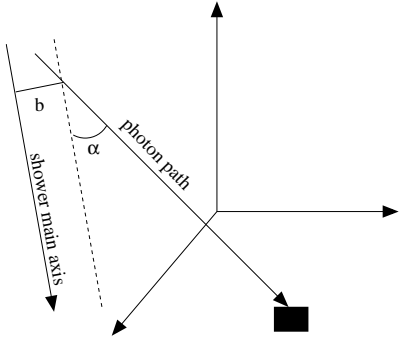}
\end{center}
\caption{Reconstruction variables}\label{fig:icrc0281_fig01}
\end{figure}

With these variables, it may be possible to both develop an iterative 3D method to geometrically reconstruct Extensive Air 
Showers seen by IACTs and effectively measure the difference between primaries to identify the primary of each event.

The separation algorithm uses the distributions of $b$ and $\alpha$ that are built with the reconstructed shower. The 
method proposed here is a preliminary one, that is still under test with simulated data.

\section{Geometrical shower reconstruction}
Geometrically, the primary particle path is a straight line defined by the direction and a position in space. The point 
chosen is located at ground level, being the point where the primary would hit if it crossed the atmosphere with no 
interaction, called the shower core position. The direction is characterized by the zenith ($\theta$) and the azimuth ($\Phi$).

The developed reconstruction method of the geometry of the EAS uses the $\alpha$, b variables discussed before. But in 
order to calculate them, it is necessary to have an approximate value of the core position and the main axis. To 
solve this problem, an approximation is originally calculated in a first stage. On the second stage, the initial guess 
is significantly improved.

As a charged particle moves through the atmosphere at a speed higher than the speed of light, it emits Cherenkov photons in a cone. 
It can be assumed that some of the photons that reached different detectors were emitted by the same particle. With a set of N 
detectors (N$>$1),the intersection of photon directions from different detectors provides a guess for points belonging to the 
shower. In reality, the directions usually do not intersect and the closest point to both is used.

The group of points obtained this way, as approximated as the algorithm is, provide useful information about the shower as 
they closely follow the actual main direction of the shower. A first value can then be computed calculating the main inertia axis. 
Furthermore, they should be positioned around the axis, and their barycentre should belong to it. The initial core position is 
simply the point at ground level, that belongs to the line defined by the computed direction and the barycentre of the set of points.

The second stage is an iterative process. In each iteration, the impact parameter is calculated for every photon. The sign of $b$ 
is related to the position of the emission point, in respect to the plane defined by the shower axis and the detector the 
photon belongs to. For each telescope, the average and standard deviation of $b$s are computed. If the reconstructed shower 
line is close to the real one, the averages should be close to zero. Here, a problem arises with the photons that reach the detector 
with directions contrary to the main axis which results in $b$ being calculated below the ground. Since this is not a valid 
solution, these points will not be used.

At the end of each iteration, the set of photons used in each detector is reduced to the ones that are close to the average, 
within a few sigma. This guarantees that in the next iteration, the new axis and core position are calculated from a group 
of points with less fluctuations. The method is iterated until the core position is stable enough (moving less than a few 
centimetres).

To test this method, a hundred $\gamma$ showers for the energies of 800 GeV, 1000 GeV, 1500 GeV, 2000 GeV and 3000 
GeV were generated. The core position was selected randomly from a square of 120 m side, while $\theta$ ranged from 0$^{\circ}$ 
to 30$^{\circ}$ and $\Phi$ from 0$^{\circ}$ to 360$^{\circ}$. To approach reality, an efficiency of 10$\%$ was imposed at the telescopes.
The average of the absolute value of the errors in the determination of the core position and the shower main axis, are 
represented in figures 2 and 3, respectively.

\begin{figure}[h]
\begin{center}
\includegraphics [width=0.4\textwidth]{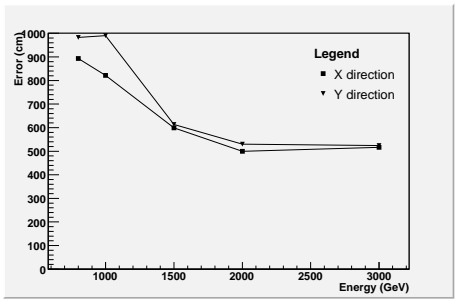}
\end{center}
\caption{Errors for the position}\label{fig:icrc0281_fig02}
\end{figure}

\begin{figure}[h]
\begin{center}
\includegraphics [width=0.4\textwidth]{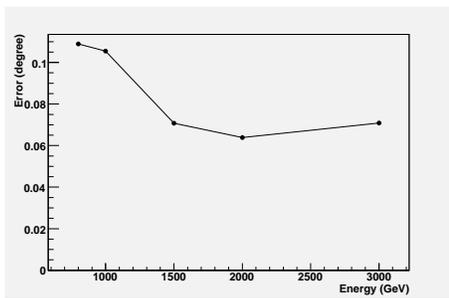}
\end{center}
\caption{Error for the deviation angle in the direction}\label{fig:icrc0281_fig03}
\end{figure}

The comparison of these results with simulated proton showers indicates that the method provides more accurate results for 
$\gamma$ induced showers than proton induced ones.

\section{Gamma/Hadron separation}
To analyse the differences, in terms of the proposed variables, between $\gamma$ induced showers and proton induced ones, one 
event of each was generated. The energy of the primary particle was 3 TeV for both cases, with a vertical direction. 
All of the Cherenkov photons emitted were used in order to build statistics. This corresponds to having an enormous detector on the ground 
with 100$\%$ efficiency.

The study of these variables was performed by analysing the distribution curves for each variable at fixed radius. For 
each distance $R$, the photons that arrived inside the corona of $R \pm 0.5$ m were considered. The radius 
presented here are 10 m, 70 m and 150 m that represent small, medium and large distances. 

The interpretation of the $\alpha$ distribution is not straightforward. It reflects not only the emission angle but, since the air 
refraction index is a function of height, also the spatial distribution of the shower electrons as well as the distribution of their 
momentum. In any case, we expect to have small $\alpha$ angles both for $\gamma$ and protons (mostly up to 1.5$^{\circ}$) but wider 
distributions for protons as the protons showers are wider.

\begin{figure}[h]
\begin{center}
\includegraphics [width=0.43\textwidth]{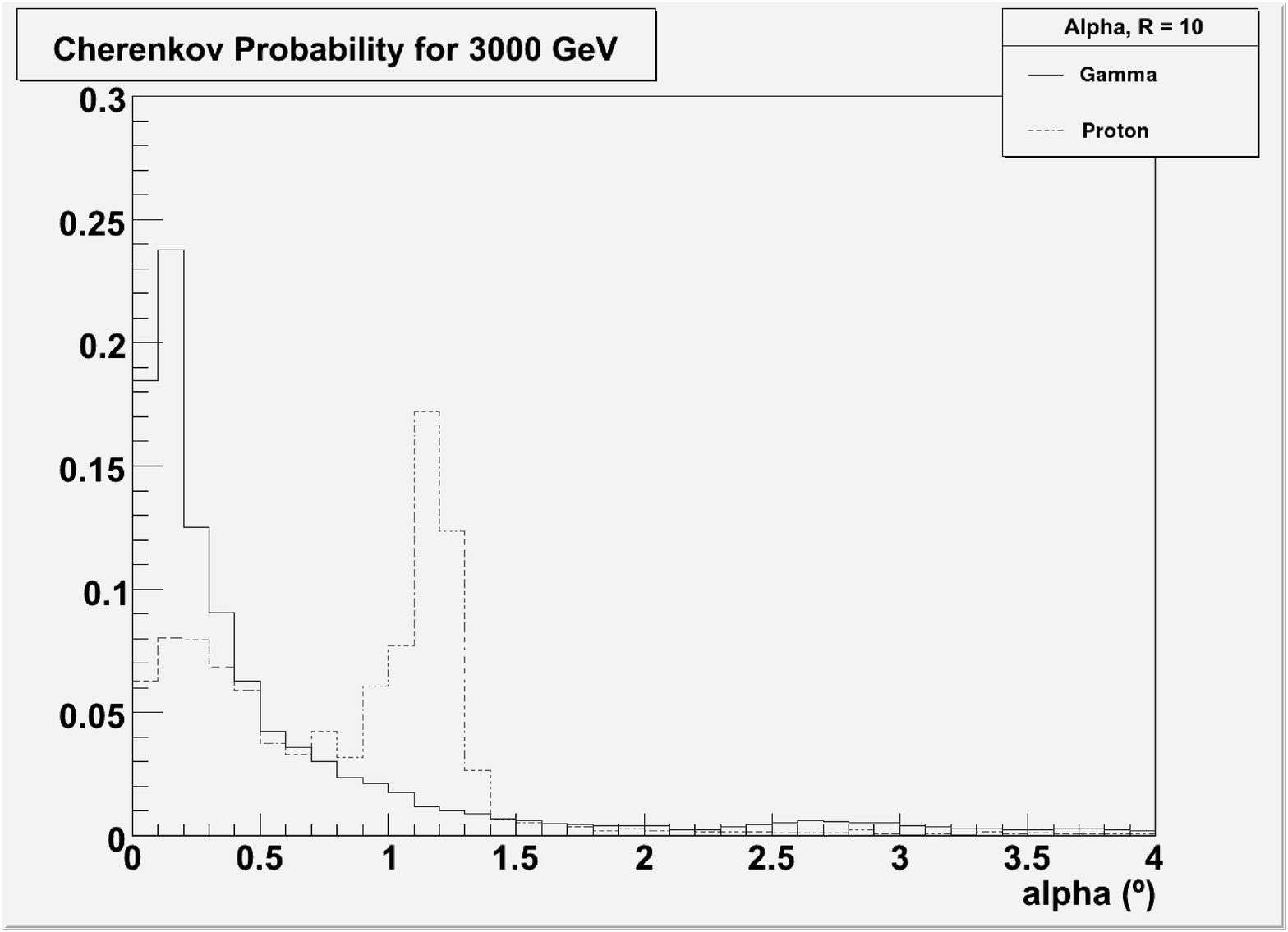}
\includegraphics [width=0.43\textwidth]{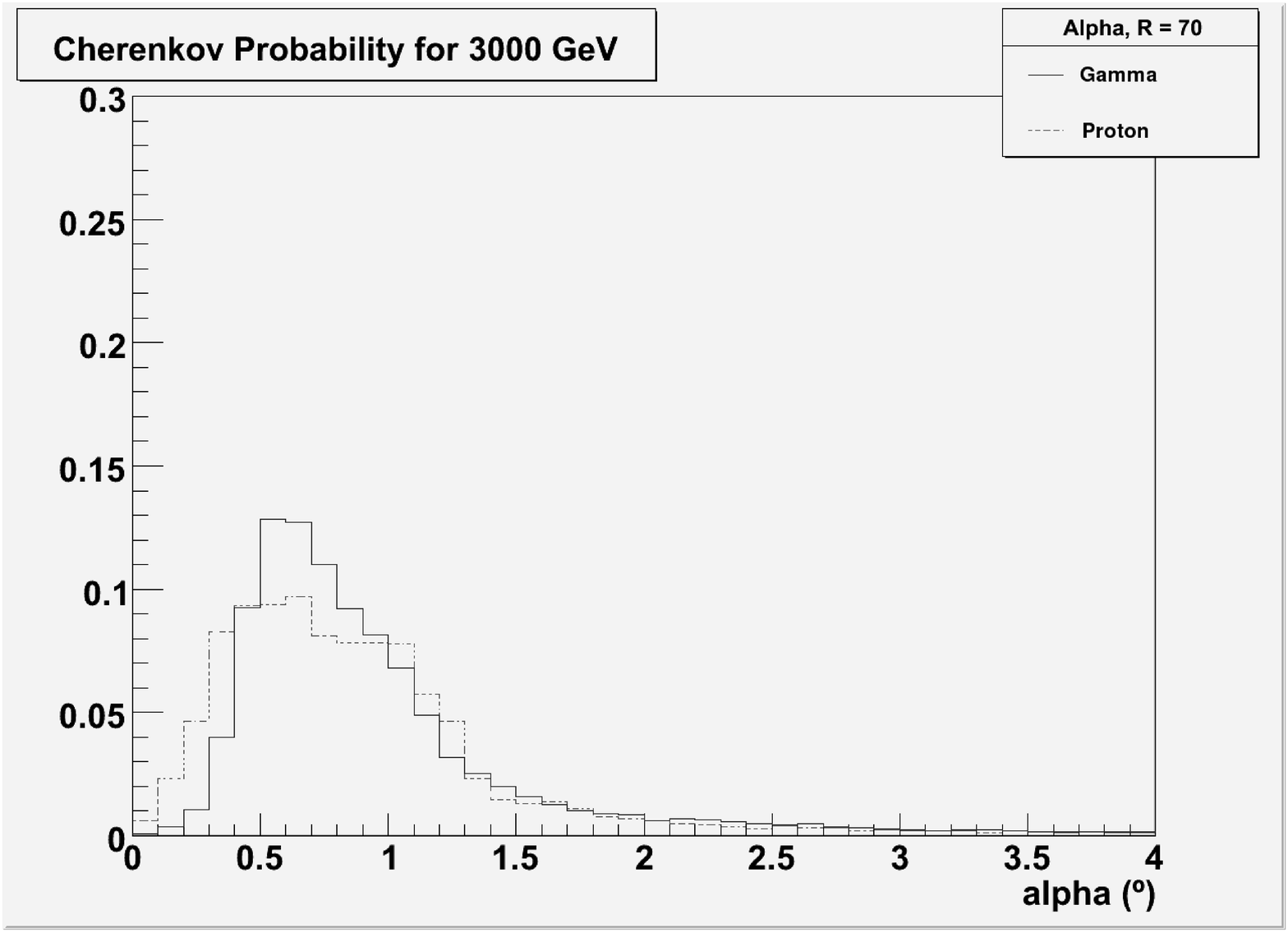}
\includegraphics [width=0.43\textwidth]{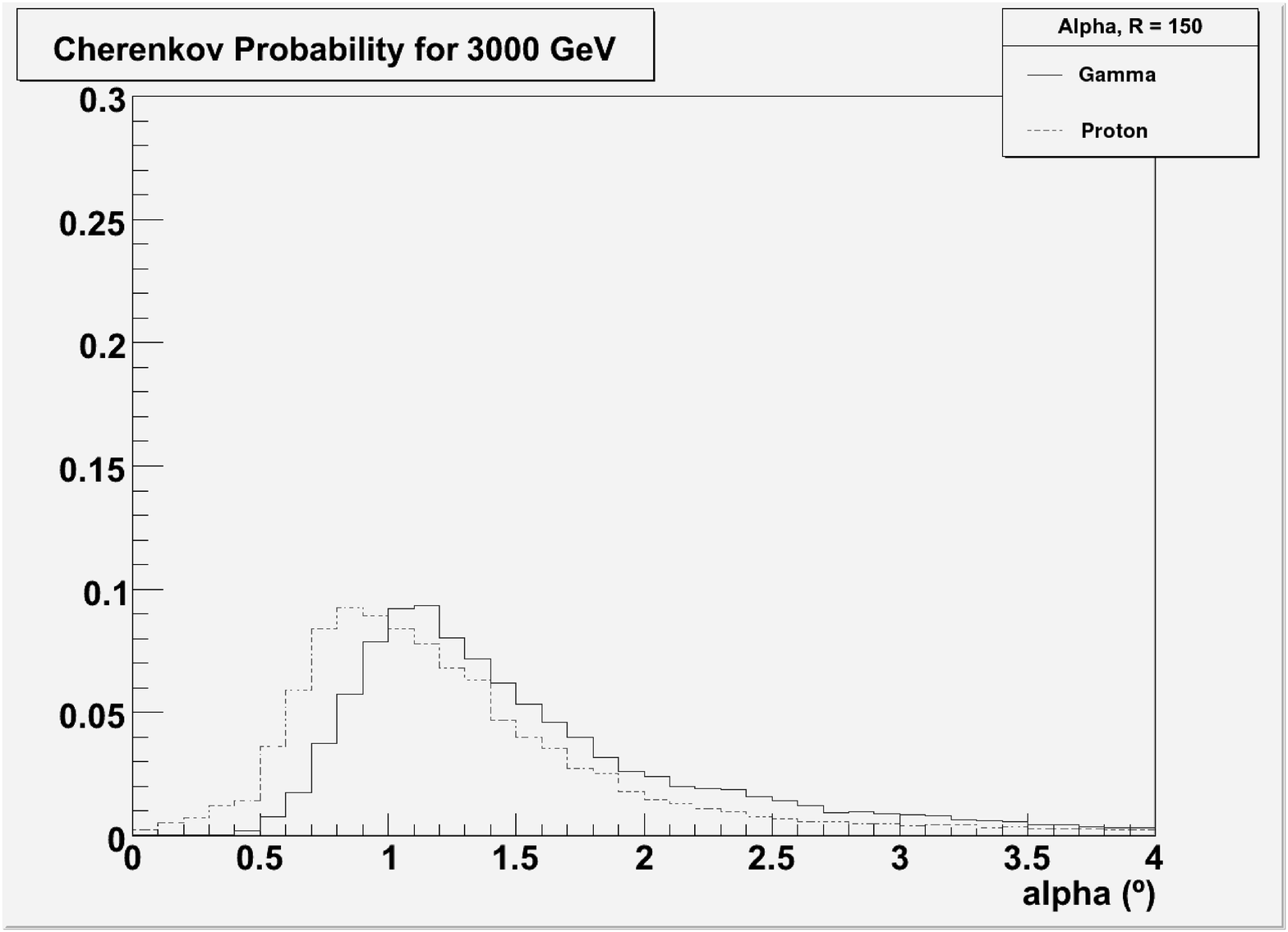}
\end{center}
\caption{$\alpha$ distribution for $R=$ 10m, $R=$ 70m, and $R=$ 150m}\label{fig:icrc0281_fig04}
\end{figure}

For $\alpha$, the distribution curves are shown in figure 4. It is easily noticed that the curves are very distinct 
at 10 m, overlapping at around 70 m, and only slightly separated at 150 m. This suggests that $\alpha$ can be used to separate the 
two types of showers when the shower is close to the telescope.

\begin{figure}[h]
\begin{center}
\includegraphics [width=0.43\textwidth]{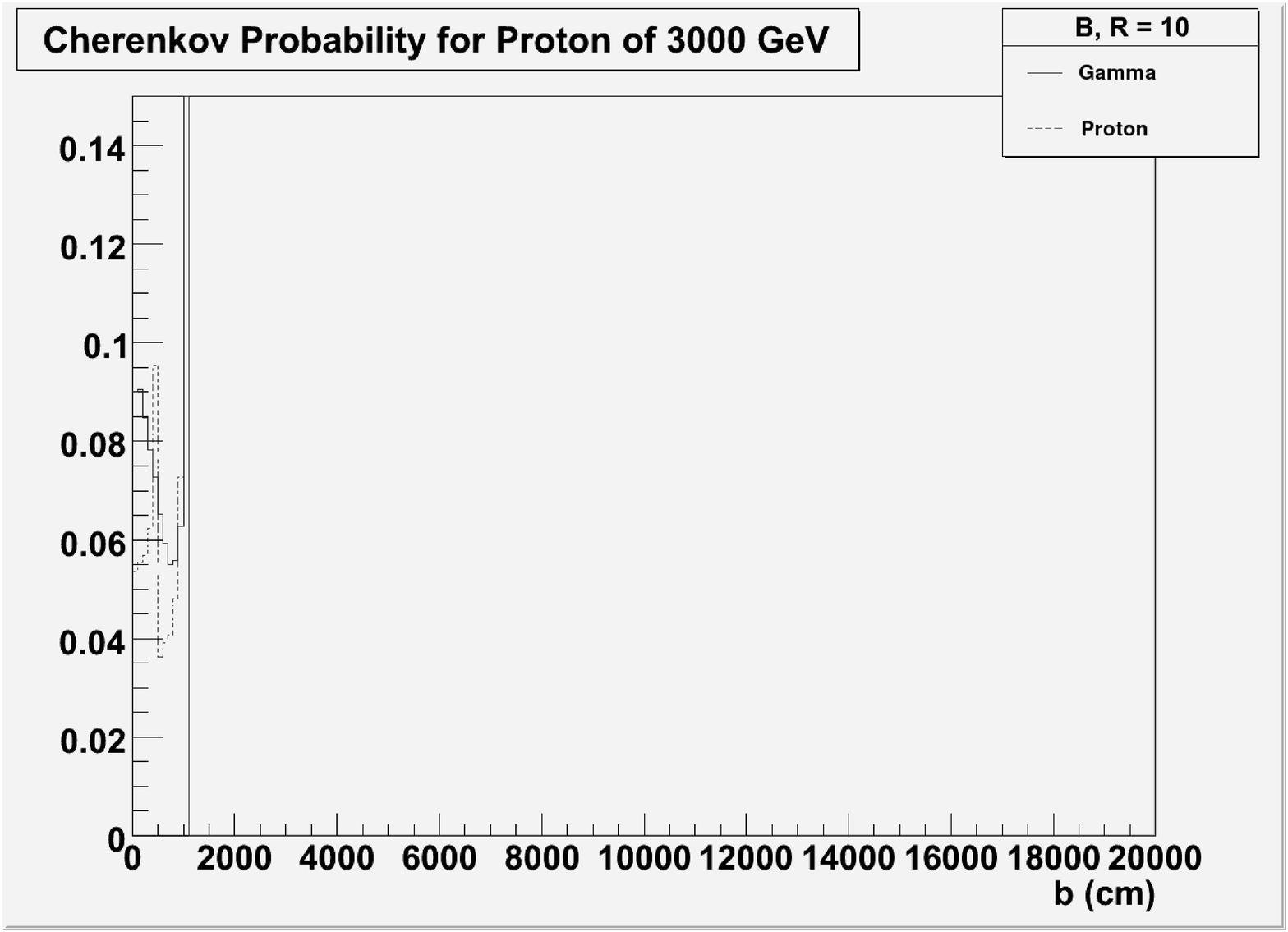}
\includegraphics [width=0.43\textwidth]{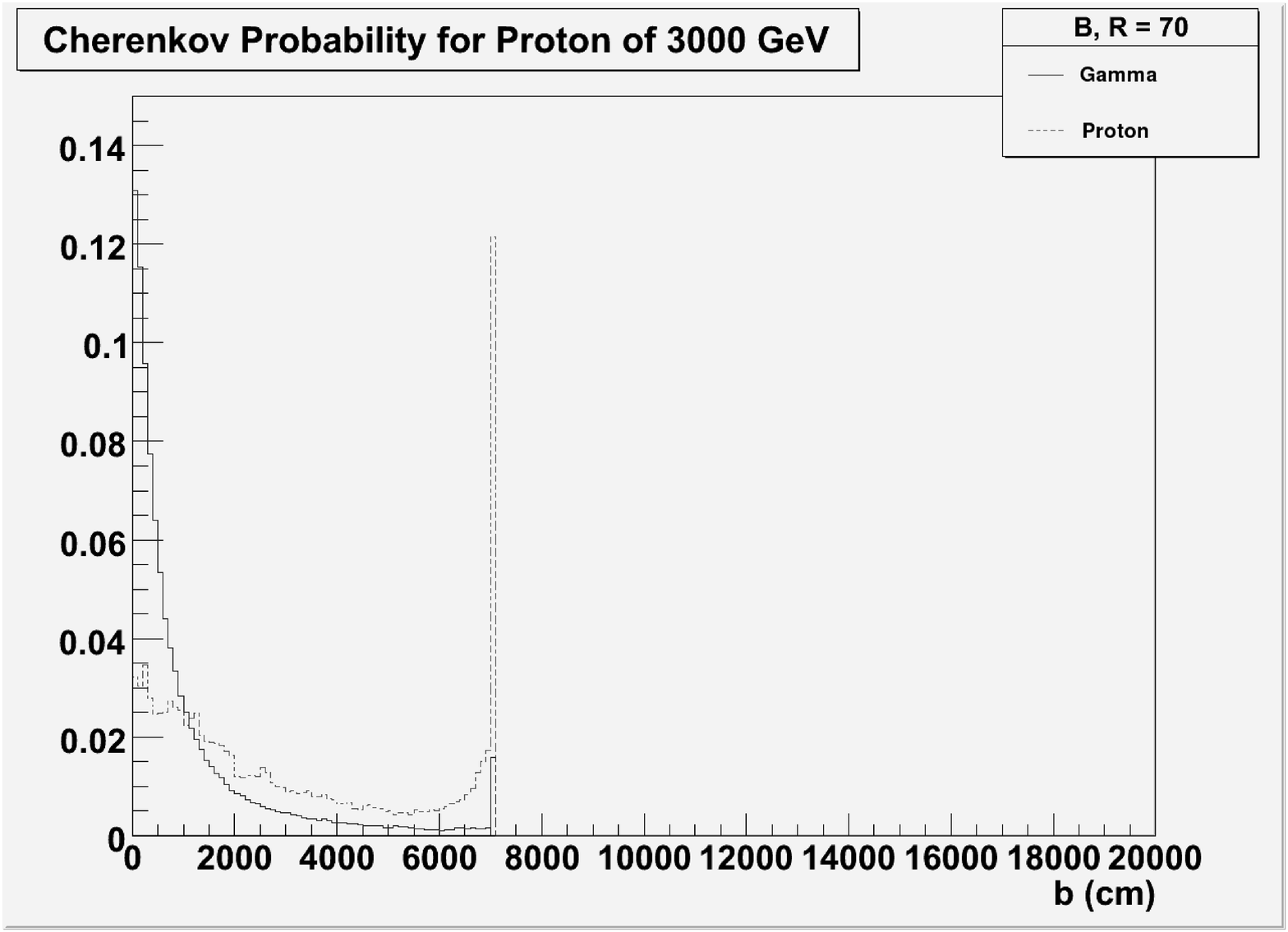}
\includegraphics [width=0.43\textwidth]{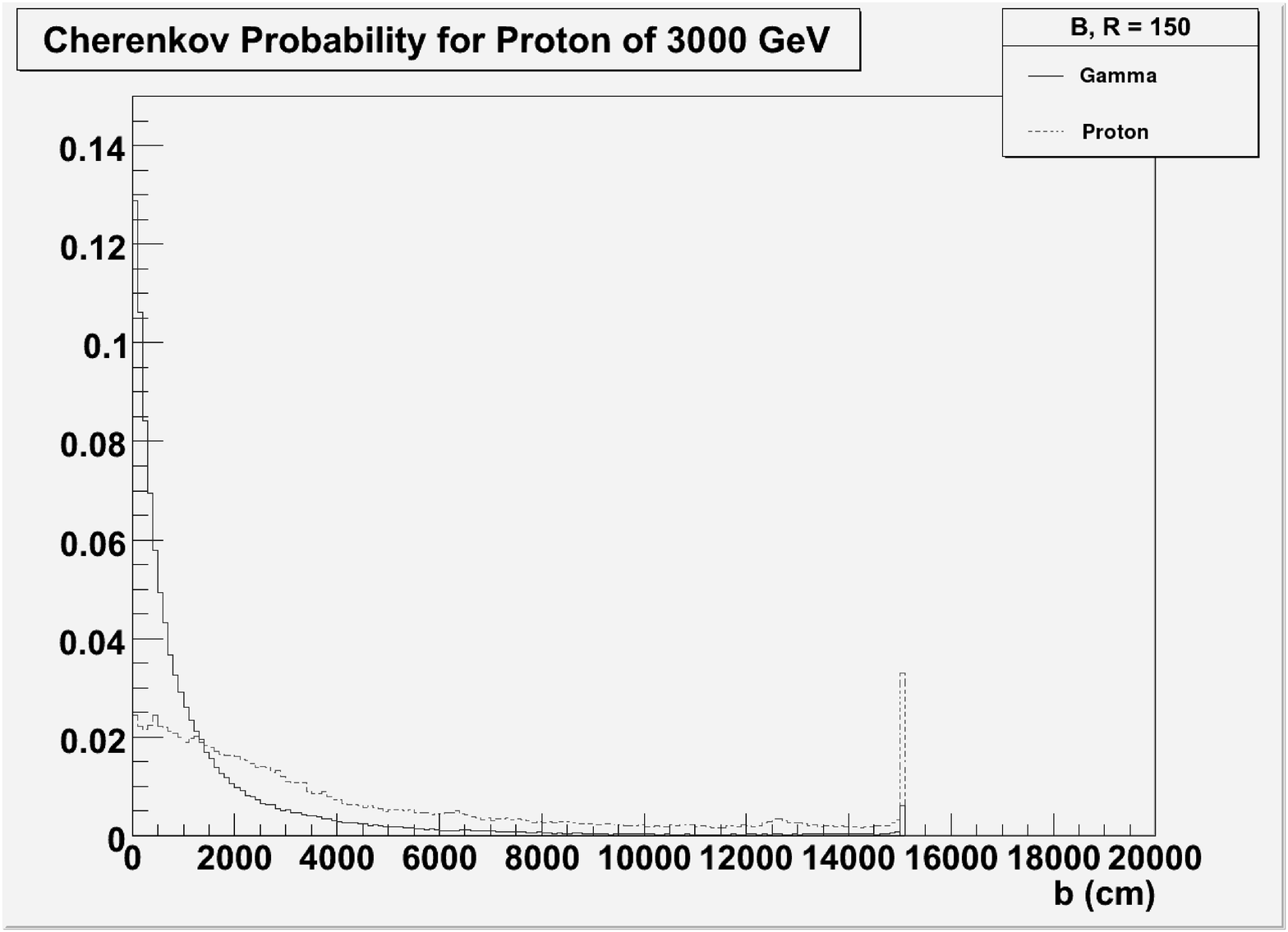}
\end{center}
\caption{Distribution of $b$ for $R=$ 10m, $R=$ 70m and $R=$ 150m}\label{fig:icrc0281_fig07}
\end{figure}

In what concerns $b$, its interpretation is simpler. It should reflect mainly the spatial distribution of the shower electrons 
and so, wider distributions are expected for protons. One should note that $b$ is set to $R$ whenever the minimum $b$ value is 
calculated at a point below ground. Then, wider showers will also be characterized by larger peaks at $b=R$. Figure 5 represents 
the $b$ distributions for the selected distances.

The analysis of the image shows that the chosen solution for the points calculated below ground introduces a "pile-up" at 
$b = R$. For small distances, the curves are almost indistinguishable as the "pile-up" dominates the 
distributions. However, for medium and high values of R the "pile-up" becomes very useful. In these cases, while $\gamma$-ray 
induced showers have almost no photons there, the proton spectrum has a well-defined peak in that region. Therefore, 
it seems that $b$ is a good variable to distinguish the two types of showers for medium and high distances.

Currently, the good estimator that combines all the considered telescopes and both variables is under construction. Clearly, 
a dependence in R has to be introduced in order to select the good variable according to the distance as each has its own region 
of influence. Although not yet complete, this path of study seems to provide a good discrimination of $\gamma$-ray induced 
showers.

\section{Conclusion and prospects}
The reconstruction method uses two spatial variables that can be understood as the emission angle and the point of emission of each 
Cherenkov photon. The tests performed so far indicate errors of about 10 m for the core position and a deviation of about 0.08$^{\circ}$ 
in the shower direction.

The proposed variables for the separation of $\gamma$ and proton showers seem to allow an efficient separation for the distances 
considered. An estimator is being built to account for the number of telescopes in use, the distance of each telescope to the 
core position and the good region for each variable.

\bibliography{icrc0281}
\bibliographystyle{plain}

\end{document}